# The γ-ray production in neutral-current neutrino-oxygen interaction in the energy range above 100 MeV

T.Mori[a], R.Yamaguchi[a], M.Sakuda[a,*], A.M.Ankowski[b,c], and O.Benhar[b]

[a]Department of Physics, Okayama University, Okayama, 700-8530, Japan
[b]INFN and Department of Physics, Sapienza  Università di Roma, I-00185 Roma, Italy
[c]Institute of Theoretical Physics, University of Wroclaw, Wroclaw, Poland

**Abstract.** We calculate the cross section of the γ-ray production from neutral-current neutrino-oxygen quasi-elastic interaction, $\nu+{}^{16}O \rightarrow \nu+p+{}^{15}N^*$, or $\nu+{}^{16}O \rightarrow \nu+n+{}^{15}O^*$, in which the residual nuclei ($^{15}N^*$ or $^{15}O^*$) lead to the γ-ray emission with $E_\gamma > 6$ MeV at the branching ratio of 41%.  Above 200 MeV, this cross section dominates over that of γ-ray production from the inelastic reaction, $\nu+{}^{16}O \rightarrow \nu+{}^{16}O^*$. In the present calculation, spectral function and the spectroscopic factors of  $1p_{1/2}, 1p_{3/2}$ and  $1s_{1/2}$  states are essential. The γ-ray production is dominated by the deexcitation of $1p_{3/2}$ state of the residual nucleus.

## OVERALL FEATURE OF NEUTRAL-CURRENT NEUTRINO-NUCLEUS INTERACTIONS FROM 10 MEV TO 1000 MEV AND THE GAMMA PRODUCTION

The γ-ray production from neutral-current neutrino-nucleus interactions at low energy ($E_\nu < 100$ MeV) was first calculated by Donnelly et al.[1,2]. Since then, many authors have calculated and discussed this process in relation to the Supernova physics and beam-dump neutrino oscillation experiments [3]. Below the neutrino energy  200 MeV, γ-rays originate from the NC inelastic reaction, $\nu+A \rightarrow \nu+A^*$,  in which the nucleus is sometimes excited to the giant resonances. Those giant resonances above particle emission threshold decay to the residual nucleus plus a nucleon, and then the excited residual nucleus sometimes emits γ-rays in the 5-10MeV region. For carbon, the giant resonances ($1^+$, 15.1 MeV and 12.7 MeV) exist below a proton or neutron emission threshold and they decay to the carbon ground state emitting γ-rays[1,2]. This process is dominant in the neutrino energy between 10 and 100 MeV region, but the cross section saturates for $E_\nu$ above 100 MeV [1,2]. The 15.1MeV γ-ray production from NC neutrino-carbon interaction from the process was measured by KARMEN experiment in the energy region from 30 to 150 MeV [3]. Just as the cross section of elastic (or coherent) neutrino-nucleus scattering ($\nu+A \rightarrow \nu+A$) saturates above 200 MeV, that of γ-ray production from NC from the inelastic reaction, ν+A→ν+A*, saturates above 200 MeV.

* sakuda@fphy.hep.okayama-u.ac.jp

For the neutrino energy $E_ν$>200MeV, the cross section of quasi-elastic interactions becomes larger than that of elatic scattering. FIG.1 shows the cross section of neutrino-oxygen elastic (coherent) scattering and quasi-elastic scattering. It is clearly shown that quasi-elastic process with a nucleon knockout dominates over the elastic process above 200 MeV. Neutrinos interact with oxygen nucleus below 100 MeV (elastic scattering), while they interact with "nucleons" inside oxygen above 200 MeV (quasi-elastic scattering). The feature of the cross section for the associated γ-ray production is similar.

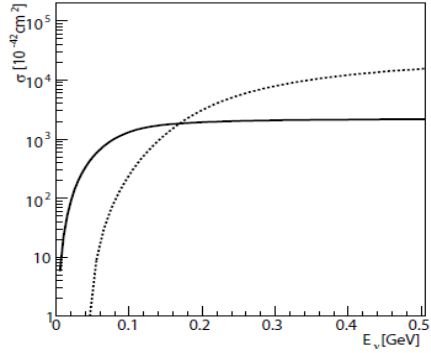

**FIGURE 1.** Feature of Neutral-Current (NC) Neutrino-Oxygen elastic (coherent) scattering (Solid line) and NC quasi-elastic scattering (Dashed line). The size of the nucleus and nucleon decides the saturation energy.

## CALUCULATION OF GAMMA-PRODUCTION IN NEUTRAL-CURRENT NEUTRINO-NUCLEUS QUASI-ELASTIC INTERACTION

We estimate for the first time the cross section of the γ-ray production from neutral-current neutrino-oxygen quasi-elastic interaction, $ν+^{16}O→ν+p+^{15}N^*$, or $ν+^{16}O→ν+n+^{15}O^*$, in which the residual nuclei ($^{15}N^*$ or $^{15}O^*$) lead to the γ-ray emission. The K2K experiment reported a preliminary result of γ-ray's from de-excitation of oxygen induced by NC neutrino-oxygen interactions in 1kton water Cherenkov detector at $<E_ν>$=1.2 GeV [4].

We calculate the cross section of the γ-ray production as,

$$\sigma_{NC}\left(ν+^{16}O→ν+p/n+X+\gamma\right)=\sigma_{NC}\left(ν+^{16}O→ν+p/n+X^*\right)\times Br\left(X^*→Y+\gamma\right), \quad (1)$$

where p and n are a proton and a neutron, X is a residual nucleus, either $^{15}N^*$ or $^{15}O^*$ including their ground states, and Y is either $^{15}N$, $^{14}N$+n, $^{14}C$+p, $^{15}O$, $^{14}N$+p, or $^{14}O$+n. FIG.2 illustrates the schematic diagrams of (a) a proton hole in $^{15}N$ induced by the reaction, $ν+^{16}O→ν+p+^{15}N^*$, and (b) the γ-ray emission from an inner shell state $1p_{3/2}$ below the nucleon emission threshold ($E_{th}$=10.2 MeV) to the ground state $^{15}N$, or that from a deep hole state $1s_{1/2}$ above $E_{th}$ through $^{14}$N*+n or $^{14}$C*+p. The γ-ray emission from the reaction, $v+^{16}O \rightarrow v+n+^{15}O^*$, is similar.

The NC quasi-elastic cross section $\sigma_{NC}(v+^{16}O \rightarrow v+p/n+X^*)$ can be calculated using the impulse-approximation and the spectral function as the charged-current quasi-elastic cross section, which was reported previously [5]. We estimate the branching ratio of γ-ray emission $Br(X^* \rightarrow Y+\gamma)$, following Ejiri's paper [6] with some improvement.

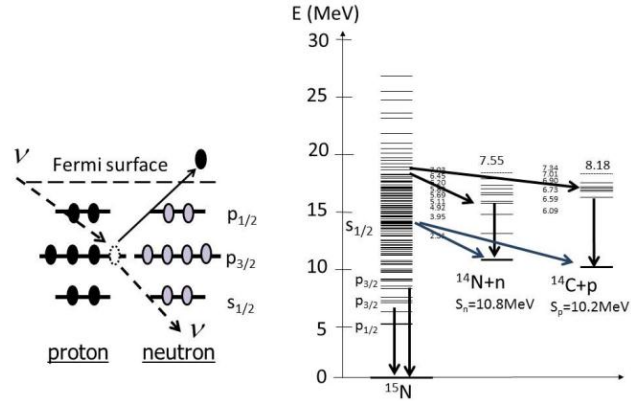

**FIGURE 2.** (a) Schematic diagram of a proton hole in $^{15}$N induced by the reaction $v+^{16}O \rightarrow v+p+^{15}N^*$, and (b) the γ-ray emission from an inner shell state $1p_{3/2}$ below the nucleon emission threshold ($E_{th}$=10.2 MeV) to the ground state $^{15}$N, or that from a deep hole state $1s_{1/2}$ above $E_{th}$ through $^{14}$N*+n or $^{14}$C*+p.

The Spectral Functions $P(p,E)$ is the probability of removing a nucleon of momentum ($p$) from ground state of the nucleus ($A$) leaving the residual nucleus ($X$) with excitation energy ($E$). FIG.3 shows the kinematics of NC quasi-elastic neutrino-oxygen reaction, $v(k)+N(p) \rightarrow v'(k')+N(p')+X$. $N(p)$ is either a proton or a neutron with momentum ($p$).

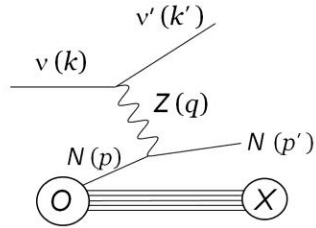

**FIGURE 3.** Kinamatic diamgram of neutral-current quasi-elastic neutrino-oxygen reaction $v(k)+N(p) \rightarrow v'(k')+N(p')+X$. $N(p)$ is either a proton or a neutron with momentum ($p$).

We stress that this definition of spectral function naturally enables us not only to calculate the NC quasi-elastic cross section, but also to estimate the probability of the residual nucleus (X) in a specific shell state, which is called a spectroscopic factor ($S_p$). This feature is essential to the present work. The spectral functions $P(p,E)$ for $^{16}O$ and $^{12}C$ are estimated by Benhar et al.[7]. The NC quasi-elastic cross section $\sigma_{NC}(\nu+^{16}O \to \nu + p/n + X^*)$ is calculated as,

$$\frac{d^2\sigma_{\nu A}}{d\Omega dE'_\nu} = \int d^3\mathbf{p} dE\, P(\mathbf{p},E) \frac{M}{p_0} \frac{E'_\nu}{E_\nu} [Z \frac{d^2\sigma_{\nu p}}{d\Omega dE'_\nu} + (A-Z) \frac{d^2\sigma_{\nu n}}{d\Omega dE'_\nu}], \quad (2)$$

where $d^2\sigma_{\nu N}/d\Omega dE'_\nu$ ($N=p,n$) is either NC $\nu p$ or $\nu n$ elastic differential cross section.

$$\frac{d^2\sigma_{\nu N}}{d\Omega dE'_\nu} = \frac{G_F^2}{16\pi^2} \frac{E'_\nu}{E_\nu} L_{\mu\nu} W^{\mu\nu} \quad (3)$$

$L_{\mu\nu}$ and $W_{\mu\nu}$ are the leptonic and hadronic tensor, respectively. We follow the notation of Ref.[8]. We use the following NC nucleon form factors $\{\tilde{F}_i^N\}$ (i=1,2),(N=p,n) and $\{\tilde{F}_A\}$ [9]:

$$\tilde{F}_i^N = \pm\frac{1}{2}(F_i^p - F_i^n) - 2\sin^2\theta_W F_i^N \quad (4)$$

$$\tilde{F}_A = \pm\frac{1}{2} F_A = \pm\frac{1}{2} \frac{1.230}{(1+Q^2/M_A^2)^2} \quad (5)$$

$\{F_i^N\}$ (i=1,2),(N=p,n) in Eq.(4) are the usual electromagnetic nucleon form factors [10] and $F_A$ is the axial form factor with $M_A$=1.10 (GeV)$^2$. (+) sign of ($\pm$) is for N=proton and (-) sign is for N=neutron in Eqs.(4-5). We obtain the dashed line of Fig.4 for the NC neutrino-oxygen quasi-elastic cross section as functions of neutrino energy. The points in closed circles are the corresponding calculation by Kolbe et al.[2]. They agree well.

Now, we estimate the braching ratio of $\gamma$-ray emission $Br(X^* \to Y+\gamma)$. When a neutrino kicks out a proton from the $^{16}O$ nucleus above the Fermi surface, a residual nucleus is left as $^{15}N$ as shown in Fig.2. In a naiive shell model, $^{16}O$ consists of $1p_{1/2}$ (ground state), $1p_{3/2}$ and $1s_{1/2}$ orbits with four, eight and four nucleons, respectively. Thus, the corresponding spectroscopic factors ($S_p$) are expected by the naiive shell model to be $S_p(1p_{1/2})$=2/8, $S_p(1p_{3/2})$=4/8, and $S_p(1s_{1/2})$=2/8, respectively. But, the important feature of the spectroscopic factors and the spectral function is that about 20% of the nucleons in $^{16}O$ do not behave like a single particle in a shell state, but they have high (p,E) components, which are highly correlated [11]. Based on O(e,e'p) data, Benhar estimated them to be $S_p(1p_{1/2})$=0.188, $S_p(1p_{3/2}$, 6.32MeV)=0.375, $S_p(1p_{3/2}$, 9.93MeV+10.7MeV)=0.06, and $S_p(1s_{1/2})$=0.188, respectively [7,12].

When the neutrino interacts with a proton in the $1p_{1/2}$ ground state, no $\gamma$-rays are produced. The $1p_{3/2}$ proton hole strength in $^{15}N$ is distributed into the 6.32 MeV, 9.93 MeV and 10.7 MeV. When the neutrino kicks out a proton in the $1p_{3/2}$ (6.32MeV) state, 6.3MeV γ-ray is produced with 100% braching ratio. In the case of a proton hole in $1p_{3/2}$ (9.93MeV) state, it emits γ-rays with energy more than 6 MeV at 100%. For $1p_{3/2}$ (10.7MeV), it decays with α particle and does not emit γ-rays [13]. In total, we estimate $Br(^{15}N(1p_{3/2}) \to ^{15}N + \gamma(>6MeV)) = 0.405$. For a deep hole state $1s_{1/2}$ above $E_{th}$, it decays through $^{14}N^*+n$ or $^{14}C^*+p$. The γ-rays are expected only when $^{14}N$ or $^{14}C$ are in the excited states. If we use TableII of Ref.[6], we estimate $Br(^{15}N \to 1s_{1/2} \to Y + \gamma(>6MeV)) = 0.027$. RCNP E148 experiment O(p,2p)N measured $Br(^{15}N(1s_{1/2}) \to Y + \gamma(>6MeV)) = 0.156 \pm 0.013$ [14]. If we use $S_p(1s_{1/2})=0.188$, then, we estimate $Br(^{15}N^* \to 1s_{1/2} \to Y + \gamma(>6MeV)) = 0.188*0.156 = 0.029$, which agrees with 0.027 above. In total, we obtain $Br(^{15}N \to Y + \gamma(>6MeV)) = 0.43$. The γ-ray production is dominated by $1p_{3/2}$ state. The contribution from $s_{1/2}$ state is very small. The γ-ray emission from the reaction $\nu+^{16}O \to \nu+n+^{15}O^*$ can be estimated in a similar way, $Br(^{15}O^* \to Y + \gamma(>6MeV)) = 0.387$ We take an average and obtain $Br(X^* \to Y + \gamma(>6MeV)) = 0.41$. The details of the calculation will be published elsewhere [12]. We estimate the cross section for $\sigma_{NC}(\nu+^{16}O \to \nu + p/n + X + \gamma)$ in the solid line in Fig.4. We also plot the points (marked with crosses and triangles) for the inelastic reaction, $\nu+^{16}O \to \nu+^{16}O^*$ with $^{16}O^* \to ^{15}N^*+p$ (x) or $^{15}O^*+n$ (▲), which are from Fig.1 of Ref.[2]. The cross sections for inelastic reaction saturate above 200 MeV[2].

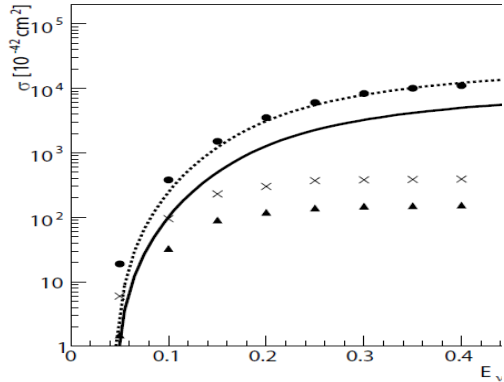

**FIGURE 4.** NC Neutrino-oxygen quasi-elastic scattering (Dashed line) and the γ-ray production. The points are from Ref.[2].

# GAMMA-PRODUCTION IN CHARGED-CURRENT NEUTRINO-NUCLEUS INTERACTIONS

The same mechanism of γ-ray production applies for CC reactions. A possibility for CC γ-ray production in neutrino-oxygen reaction, $\nu_e+^{16}O \to e^-+^{16}F^*$ and $\bar{\nu}_e+^{16}O \to e^++^{16}N^*(0^-,3^-,1^-)$, for E<100 MeV, was pointed out by Haxton [15] and that for CC neutrino-carbon interactions was discussed by Kolbe et al.[16]. Again the cross sections from the reactions saturate above 200 MeV. We also point out that for $E_\nu>200$ MeV the γ-ray production for CC neutrino-nucleus quasi-elastic interactions, $\nu_e + {}^{16}O \rightarrow e^- + p + {}^{15}O^*$ and $\bar{\nu}_e + {}^{16}O \rightarrow e^+ + n + {}^{15}N^*$, can happen by the same mechanism which we explained in the present article. We already know that ${}^{15}N^*$ and ${}^{15}O^*$ give γ-rays at a branching ration of about 40%. So, we predict that CC quasi-elastic (1N-knockout) events should have γ-ray (>6MeV) at about 40%.

## SUMMARY

The γ-ray production from excited nuclei in NC neutrino-nucleus interactions plays an important role in Supernova Physics and neutrino-oscillation experiments. We have estimated the cross section of γ-production from neutrino-oxygen quasi-elastic interaction using the spectral function. We find that the cross section is larger at E>200MeV than that of γ-production originating from the excitation of giant resonances of ${}^{16}O$. The γ-rays (>6 MeV) from neutrino-oxygen quasi-elastic interaction are contributed most from $1p_{3/2}$ state of ${}^{15}N$ and ${}^{15}O$ and the braching ratio is about 40%. We note that CC quasi-elastic process will produce γ-rays (>6MeV) at about 40% by the same mechanism.

## ACKNOWLEDGMENTS

We thank Profs. T.Sato, M.Yosoi and A.Tamii (Osaka) for useful discussions. This work is supported in part by the Grant-In-Aid for the Japan Society for Promotion of Science (No.23340073 and No.21224004). A.M.A. was supported by the Polish Ministry of Science and Higher Education under Grant No. 550/MOB/2009/0.